\documentclass[preprint,pra,showpacs,showkeys]{revtex4}
\usepackage{graphicx,amsmath,amsfonts,amssymb}
\usepackage{times}

\begin{document}

\title{The entanglement in one-dimensional random XY spin
chain with Dzyaloshinskii-Moriya interaction \footnote{Supported by
the Key Higher Education Programme of Hubei Province under Grant No
Z20052201, the Natural Science Foundation of Hubei Province, China
under Grant No 2006ABA055, and the Postgraduate Programme of Hubei
Normal University under Grant No 2007D20.}}

\author{C. J. Shan\footnote{ E-mail: scj1122@163.com}}
\author{W. W. Cheng}
\author{T. K. Liu\footnote{Corresponding author. E-mail:
tkliuhs@163.com}}
\author{Y. X. Huang}
\author{H. Li}
\affiliation{College of Physics and Electronic Science, Hubei Normal
University, Huangshi 435002, China}
\date{\today}

\begin{abstract}

The impurities of exchange couplings, external magnetic fields and
Dzyaloshinskii--Moriya (DM) interaction considered as Gaussian
distribution, the entanglement in one-dimensional  random $XY$ spin
systems is investigated by the method of solving the different
spin-spin correlation functions and the average magnetization per
spin. The entanglement dynamics at central locations of
ferromagnetic and antiferromagnetic chains have been studied by
varying the three impurities and the strength of DM interaction. (i)
For ferromagnetic spin chain, the weak DM interaction can improve
the amount of entanglement to a large value, and the impurities have
the opposite effect on the entanglement below and above critical DM
interaction. (ii) For antiferromagnetic spin chain, DM interaction
can enhance the entanglement to a steady value. Our results imply
that DM interaction strength, the impurity and exchange couplings
(or magnetic field) play competing roles in enhancing quantum
entanglement.
\end{abstract}



\maketitle

\section{Introduction}
Entanglement is not only  important in the quantum information
processing (QIP), such as quantum teleportation,$^{[1]}$ dense
coding,$^{[2]}$ quantum secret sharing,$^{[3]}$ quantum
computation,$^{[4]}$ but also relevant to quantum phase
transitions$^{[5]}$ in condensed matter physics. In order to realize
quantum information process, great effort has been devoted to
studying and characterizing the entanglement in cavity
QED.$^{[6-8]}$ Now, much attention has been paid to the entanglement
in spin systems, such as the Ising model$^{[9]}$ and all the kinds
of Heisenberg XY XXZ XYZ models.$^{[10-13]}$ However; as far as we
know, most discussions mentioned above merely focused on the models
with spin exchange couplings, while Dzyaloshinskii--Moriya
interaction has seldom been taken into account. The antisymmetric DM
interaction, introduced by Dzyaloshinskii and Moriya, is a
combination of superexchange and spin-orbital interactions. In fact,
some one-dimensional and two-dimensional spin models  have
manifested such interactions.$^{[14,15]}$ Therefore, it is
worthwhile including DM interaction in the studies of spin chain
entanglement.
\\\indent
Impurities necessarily exist in real materials and their effects are
more pronounced in condensed matter physics. Thus, it is important
to study the effects of impurities in view of the possible
realizations of one-dimensional ferromagnetic and antiferromagnetic
chains. In the previous researches, the impurity effects on the
quantum entanglement have been studied in a three-spin
system$^{[16,17]}$ and a large spin systems under zero
temperature.$^{[18]}$ However, in these works, they have just
studied single impurity.
\\\indent
Recently, Huang {\it et al.},$^{[19]}$ Osenda {\it et al.}$^{[20]}$
and we$^{[21]}$ have demonstrated that for a class of
one-dimensional magnetic systems, entanglement can be controlled and
tuned  by introducing impurities into the systems. For the pure
case, Osterloh {\it et al.}$^{[22]}$ examined the entanglement
between two spins of position $i$ and $j$ in the spin chain as the
system goes through quantum phase transition. They demonstrated that
entanglement shows scaling behaviour in the vicinity of the
transition point.  For a two-qubit spin chain with DM interaction,
researchers$^{[23,24]}$ have considered thermal entanglement and
teleportation. For a particular spin system the allowed components
of the DM interaction are determined by the corrections to the
energy symmetry of the spin complex.  Since the DM terms break
spin-spin rotational symmetry, we need to calculate how spin
exchange couplings and DM interaction have effect on the
entanglement and phase transition point. It is an interesting
quantum phenomenon that the entanglement shares many features with
quantum phase transition (QPT), QPT is a critical change in the
properties of the ground state of a many body system due to
modifications in the interactions among its constituents. The
associated level crossings lead to the presence of non-analyticities
in the energy spectrum. Therefore, the knowledge about the
entanglement, the non-local correlation in quantum systems, is
considered as the key to understand QPT. That is the purpose and
motivation of the present work to investigate the behaviour of
entanglement at and around the quantum critical point in
one-dimensional $XY$ spin system with DM interaction, which can
display a variety of interesting physical phenomena providing new
insight in  two-site entanglement and the related QPT as well under
the effect of the impurities of exchange couplings, external
magnetic fields and DM interaction.
\\\indent
We consider Heisenberg $XY$ model of $N$ spin-$\frac{1}{2}$
particles with nearest-neighbour interactions. In the presence of
impurities and DM interaction,$^{[25]}$ one-dimensional Hamiltonian
is given by$^{[19]}$
\begin{eqnarray}
H=-\frac{1+\gamma }{2}\sum_{i=1}^{N}J_{i,i+1}\sigma _{i}^{x}\sigma
_{i+1}^{x}-\frac{1-\gamma
}{2}\times\sum_{i=1}^{N}J_{i,i+1}\nonumber\\\times\sigma
_{i}^{y}\sigma _{i+1}^{y}-\sum_{i=1}^{N}h_{i}\sigma
_{i}^{z}-\frac{1}{2}\sum_{i=1}^{N}\vec{D}_{i,i+1}\cdot(\vec{\sigma}
_{i}\times\vec{\sigma}_{i+1})
\end{eqnarray}
where $J_{i,i+1}$ and $D_{i,i+1}$ are exchange interaction and DM
interaction along $z$-direction between sites $i$ and $i+1$
respectively, $h_{i}$ is the strength of external magnetic field on
site $i$, $\sigma^{x,y,z}$ are the Pauli matrices, $\gamma$ is the
degree of anisotropy and $N$ is the number of sites. For all the
interval $0<\gamma\leq1$ and $N=\infty$, they undergo a quantum
phase transition at the critical value $\lambda_{c}=1$. The periodic
boundary conditions satisfy $\sigma _{N+1}^{x}=\sigma _{1}^{x},
\sigma _{N+1}^{y}=\sigma _{1}^{y}, \sigma _{N+1}^{z}=\sigma
_{1}^{z}$. Let us define the raising and lowing operators
$a_{i}^{+}$, $a_{i}^{-}$ and introduce Fermi operators $c_{j}^{+}$
and $c_{j}$,$^{[26]}$ the Hamiltonian has the form
\begin{eqnarray}
H=-\sum_{i=1}^{N}[((J_{i,i+1}+iD_{i,i+1})c_{i}^{+}c_{i+1}+h.c)+\nonumber\\(J_{i,i+1}\gamma
c_{i}^{+}c_{i+1}^{+}+h.c)]-2\sum_{i=1}^{N}h_{i}(c_{i}^{+}c_{i}-\frac{
1}{2})
\end{eqnarray}
In this study,  the exchange interaction has the form
$J_{i,i+1}=J(1+\alpha_{i,i+1})$, where $\alpha$ introduces the
impurity in a Gaussian form centered at $\dfrac{N+1}{2}$ with
strength or height $\zeta$, $\alpha _{i,i+1}=\zeta\exp\Big(-\epsilon
\Big(i-\dfrac{N+1}{2}\Big)\Big)$, $\epsilon$ is the value of the
width of the distribution. For $J<0$, the spin chain is
antiferromagnetic; for $J>0$, the spin chain is ferromagnetic. The
external magnetic field and Dzyaloshinskii--Moriya interaction take
the form $h_{i}=h(1+\beta _{i})$ and
$D_{i,i+1}=D(1+\eta_{i,i+1})\exp\Big(\dfrac{\pi}{2}i\Big)$, where
$\beta _{i}=\xi \exp\Big(-\epsilon\Big(i-\dfrac{N+1}{2}\Big)\Big)$,
$\eta_{i,i+1}=\kappa\exp\Big(-\epsilon\Big(i-\dfrac{N+1}{2}\Big)\Big)$.
When $\alpha=\beta=\eta=0$, we recover the pure case; when $\eta=0$,
we recover the case described in Ref.\,[19]. For the distributions
of exchange interaction impurity, Dzyaloshinskii--Moriya interaction
impurity and the magnetic field impurity, we fix the value of width
of the distribution at $\epsilon=0.1$ in all the calculations.  As
the center $\Big(\dfrac{N+1}{2}\Big)$ and the width ($\epsilon$) of
the Gaussian distribution are fixed, we can obtain different
impurities $\alpha,\, \beta,\, \eta$ of the Gaussian distributions
only by changing strengths or heights $\zeta,\, \xi,\, \kappa$. By
introducing the dimensionless parameter $\lambda=J/2h$, the
symmetrical matrix $A$ and the antisymmetrical $B$, the Hamiltonian
becomes
\begin{eqnarray}
H=\sum_{i,j=1}^{N}[c_{i}^{+}A_{i,j}c_{j}+\frac{1}{2}%
(c_{i}^{+}B_{i,j}c_{j}^{+}+h.c)]
\end{eqnarray}
The above Hamiltonian can be diagonalized by making linear
transformation of the fermionic operators $\eta
_{k}=\sum_{i=1}^{N}g_{ki}c_{i}+h_{ki}c_{i}^{+},\eta
_{k}^{+}=\sum_{i=1}^{N}g_{ki}c_{i}^{+}+h_{ki}c_{i},$ then the
Hamiltonian becomes
\begin{eqnarray}
H=\sum_{k=1}^{N}\Lambda _{k}\eta _{k}^{+}\eta _{k}+const,
\end{eqnarray}
two coupled  matrix equations satisfy $\phi _{k}(A-B)=\Lambda
_{k}\psi _{k},$ $\psi _{k}(A+B)=\Lambda _{k}\phi _{k},$ where the
components of the two column vectors $\phi _{ki},\psi _{ki}$ are
given by $\phi _{ki}=g_{ki}+h_{ki},\psi _{ki}=g_{ki}-h_{ki}.$
Finally, the ground state of the system $|\psi_{0}\rangle $ can be
written as $\eta_{k}|\psi_{0}|=0$.

Using Wick's theorem,$^{[27]}$ spin-spin correlation functions for
the ground state and the average magnetization per spin can be
expressed as\\
 $S_{lm}^{x}=\frac{1}{4}\left(
\begin{array}{cccc}
G_{l,l+1} & G_{l,l+2} & \cdots  & G_{l,m} \\
\vdots  & \vdots  & \ddots  & \vdots  \\
G_{m-1,l+1} & G_{m-1,l+1} & \cdots  & G_{m-1,m}%
\end{array}%
\right)$,\\
$S_{lm}^{y}=\frac{1}{4}\left(
\begin{array}{cccc}
G_{l+1,l} & G_{l+1,l+1} & \cdots  & G_{l+1,m-1} \\
\vdots  & \vdots  & \ddots  & \vdots  \\
G_{m,l} & G_{m,l+1} & \cdots  & G_{m,m-1}%
\end{array}%
\right)$,\\
$S_{lm}^{z}=\frac{1}{4}(G_{l,l}G_{m,m}-G_{m,l}G_{l,m}),M_{i}^{z}=\frac{1}{2}%
G_{i,i}$ \\
where $G_{i,j}=-\sum_{k}^{N}\psi _{ki}\phi _{kj}$ Next, we give the
expression of concurrence that quantifies the amount of entanglement
between two qubits.

For a system described by the density matrix $\rho$, the concurrence
$C$ reads$^{[28]}$
\begin{eqnarray}
C(\rho )=\max (0,\lambda _{1}-\lambda _{2}-\lambda _{3}-\lambda
_{4})
\end{eqnarray}
where $\lambda_{1}$, $\lambda_{2}$, $\lambda_{3}$, and $\lambda_{4}$
are the eigenvalues (with $\lambda_{1}$ being the largest one) of
the spin-flipped density operator $R$, which is defined by
$R=\sqrt{\sqrt{\rho }\tilde{\rho} \sqrt{\rho }}$, where
$\tilde{\rho} =(\sigma_{y}\otimes \sigma_{y})\rho^*(\sigma
_{y}\otimes \sigma_{y})$; $\tilde{\rho}$ denotes the complex
conjugate of $\rho$; $\sigma_{y}$ is  the usual Pauli matrix. Using
the operator expansion for the density matrix and the symmetries of
the Hamiltonian,$^{[29]}$ in the basis states $\{|\uparrow \uparrow
\rangle,\, |\uparrow \downarrow \rangle,\, |\downarrow
\uparrow\rangle, |\downarrow \downarrow \rangle\}$, $\rho$ has the
general form. We can express all the matrix elements in the density
matrix in terms of different spin-spin correlation functions.\\
\indent In this study, we focus our discussion on the transverse
Ising model with $\gamma=1$. We examine the dynamics of entanglement
in varying the impurities of exchange couplings, external magnetic
fields  and Dzyaloshinskii--Moriya interaction. First, we examine
the change of the entanglement for the nearest-neighbouring
concurrence $C(i,i+1)$ for different values of the impurity as the
parameter $\lambda$ varies.  Figure 1 depicts the
nearest-neighbouring concurrence $C(49,50)$ as a function of the
reduced coupling constant $\lambda$ at different values of the
exchange couplings impurity $\zeta$ and external magnetic fields
impurity $\xi$ with the system size $N=99$. Figure 1(a) shows the
change of concurrence $C(49,50)$ as a function of $\lambda$ for
different values of exchange couplings impurity with $D=0$, i.e. in
the absence of DM interaction. For the case of $\lambda>0$, we can
see that the concurrence increases and arrives at a maximum close to
the critical point $\lambda_{c}$, while it is close to zero above
$\lambda_{c}$. As $\zeta$ increases the concurrence tends to
increase faster, and $\lambda_{m}$, where concurrence approaches a
maximum, shifts to left very rapidly. This is consistent with the
result reported in Refs.\,[19,21] (Fig.1). A similar behaviour can
be seen for the case of $\lambda<0$, that is to say, the
entanglement has equal value for ferromagnetic and antiferromagnetic
chains with the same $|\lambda|$. The effect of the external
magnetic field $\xi$ in the Gaussian distribution is also shown in
Fig.1(b). However, different from the effect of the exchange
couplings, the concurrence increases slowly and tends to moving to
infinity by increasing the value of the parameter $\xi$. This is
also consistent with the result in Refs.[19,21] (Fig.1). In
Figs.1(c) and 1(d), taking DM interaction into account, we give a
plot of the concurrence against exchange couplings impurity and
external magnetic field impurity with $D=0.5|J|$. As $\zeta$
increases, the concurrence increases slowly and the peak value
decreases, which can be seen in Fig.1(c), different from the result
in Fig.1(a) for ferromagnetic spin chain. Moreover, some interesting
physical phenomena occur for the antiferromagnetic chain, for
example, the concurrence decreases to zero at the critical point
($\lambda_{0}$) and increases from zero to a finite steady value
across the transition point. Therefore, we can further understand
the relation between the entanglement and quantum transition. In
Fig.1(d), the numerical calculations show that the steady
concurrence decreases with the increase of $\xi$ for the
antiferromagnetic chain, which indicates that the behaviour is very
different from those in Fig.1(b). Now weexplain why the curves of
concurrence have some maximum or minimum at some special values of
the DM interactions and external magnetic fields. As $[\partial
C(49,50)/\partial \lambda]_{\lambda_{c}}$ diverges, the maximal or
minimal entanglement will not occur at the critical point but in the
vicinity of the transition point ($[\partial C(49,50)/\partial
\lambda]_{\lambda_{m}}=0,$ $[\partial^{2} C(49,50)/\partial^{2}
\lambda]_{\lambda_{m}}<0$ or $[\partial^{2}C(49,50)/\partial^{2}
\lambda]_{\lambda_{m}}>0$). In our model, when $D=0$, quantum
transition point
$\lambda_{c}=\dfrac{1+\beta_{i,i+1}}{1+\alpha_{i,i+1}}$, from the
above expression, we can know the transition point shifts and is
affected by the impurities of exchange couplings and external
magnetic fields. For the entanglement length (or the correlation
length), the position of the related maximal and minimal concurrence
will shift in the same way. However, DM interactions lead to
different coefficients in the first two  parts of Eq.(2), so the
critical point $\lambda_{c}$ occurs between the two ones
$\Big(\dfrac{1+\beta_{i,i+1}}{1+\alpha_{i,i+1}},
\dfrac{1+\beta_{i,i+1}}{1+\alpha_{i,i+1}-\frac{D}{J}(1+\eta_{i,i+1})}\Big)$.
For the case of $J<0$, there exist two critical points.  Of course,
we can figure out exact critical value and maximum or minimum of the
concurrence through solving the first order and second order
derivative  of the entanglement respectively.
\\\indent
From Fig.1, we can see that DM interaction plays an important role
in enhancing entanglement, so it is necessary to study the effect of
DM interaction on the entanglement. In Fig.2, we show the results of
the nearest-neighbouring concurrence as a function of the parameter
$\lambda$ for DM interaction impurity at different strengths of DM
interaction $D$. We can easily find that the competing roles played
by DM interaction impurity $\kappa$, strength $D$ and exchange
couplings $J$ (the external magnetic field is fixed) in enhancing
quantum entanglement will exist in spin chain. The competing effect
leads to shift of the critical point and the entanglement. The
results show that when the absolute value of $\lambda$ is below
$\lambda_{c}$, the concurrence only increases with $|\lambda|$,  DM
interaction impurities will have no effect on the entanglement once
strength is fixed, i.e. exchange couplings is predominant in the
competing role. The effect of weak DM interaction strength
$D=0.1|J|$ is shown in Fig.2(a). Contrast to the exchange couplings
impurity, DM interaction impurity can enhance the entanglement, the
concurrence increases and tends to move to infinity($>0$) by
increasing the value of the parameter $\kappa$. It is interesting to
find that the entanglement peak and steady value between the nearest
neighbours with $D=0.5|J|$ increase to a value larger than those in
Fig.\,2(a). With the increasing $D$, in Fig.2(c), the concurrence
decreases as $\kappa$ increases. We can imagine that there must be a
critical DM strength ($D_{c}$), below $D_{c}$, impurity enhances
entanglement, while above $D_{c}$, impurity shrinks entanglement. In
other words, at some special values of the DM interactions, the
entanglement varies at different critical vicinities, which is
similar to the analysis in Fig.1. The comparison among the different
curve in Fig.2(d) shows that the concurrence decreases rapidly above
$\lambda_{c}$ by increasing the value of the parameter $\kappa$,
which is different from the results obtained from Figs.\,2(a) and
2(b). That is to say, the strong $D$ is not helpful to keeping the
better entanglement for Gaussian distribution.\\
\indent The effect of  DM strength is demonstrated in Fig.3 by the
evolutions of the concurrence. Figure 3(a) corresponds to the case
of $\kappa=0$, the nearest-neighbouring concurrence increases with
the increase of $D$, a critical point occurs with small DM
interaction strength and the peak of the maximal entanglement
becomes larger. It is the DM interaction that leads to considerable
different evolutions of the entanglement, hence the entanglement is
rather sensitive to any small change with the DM interaction. Thus,
by adjusting DM interaction one can obtain a strong entanglement.
Similar behaviours to those in Figs.2(c)and 2(d) are shown in
Figs.3(c) and 3(d), we can see that DM interaction strength is not
certain to enhance the entanglement, and the entanglement tends to
be reduced in the presence of strong DM interaction at $\kappa=1.0$.
The results we have obtained here are also consistent with those in
Fig.2. \\
\indent According to finite-size scaling analysis, the two-site
entanglement is considered as a function of the system size
(including the thermodynamic limit) and the distance
$|\lambda-\lambda_{c}|$ from the critical point. The entanglement
can approximately collapse to a single curve for different system
sizes ranging from 41 up to 401, thus all key ingredients of the
finite-size scaling are present in the concurrence. The first order
derivative around the critical point becomes sharper (the peak
position $\lambda_{\min}$ approaches the critical point
$\lambda_{c}$) as the system size increases, and is expected to be
divergent in an infinite system ($\partial C/\partial \lambda=A_{1}
\ln|\lambda-\lambda_{c}| +{\rm const}$). Though there is no
divergence when $N$ is finite, the anomalies are obvious. Its value
diverges logarithmically with the increasing system size as
$\partial C/\partial \lambda=A_{2}\ln N +{\rm const}$. Thus, we can
see that the QPT of the system is reflected by the behaviour of the
concurrence and its $\lambda$ derivative and finite size scaling is
fulfilled over a very broad range of values of $N$, which are of
interest in quantum information.
\\\indent
In summary, from the above analysis, it is clearly noted that the
three different impurities and DM interaction strength, which play
the competing roles in enhancing quantum entanglement, have a
notable influence on the nearest-neighbouring concurrence in the
one-dimensional $s=\dfrac{1}{2}$ random $XY$ spin system. The
nearest-neighbouring concurrence exhibits some interesting
phenomena. For an antiferromagnetic spin chain, there is a critical
point where the entanglement is zero. DM interaction is predominant
in the competing role and can enhance the entanglement to a steady
value. For a ferromagnetic spin chain, the weak DM interaction can
improve the amount of entanglement to a large value. However, under
condition of strong DM interaction, there is a critical point
$D_{c}$ where  the impurities have the opposite effect on the
entanglement below and above $D_{c}$. Thus we can employ DM
interaction strength as well as three different impurities to
realize quantum entanglement control.  For the case of $\gamma\neq1$
(XY model) or the next nearest-neighbouring concurrence related QPT,
we will present further reports in the future.\\

\newpage
\begin{figure}
\begin{center}
\includegraphics[width=1.0\textwidth]{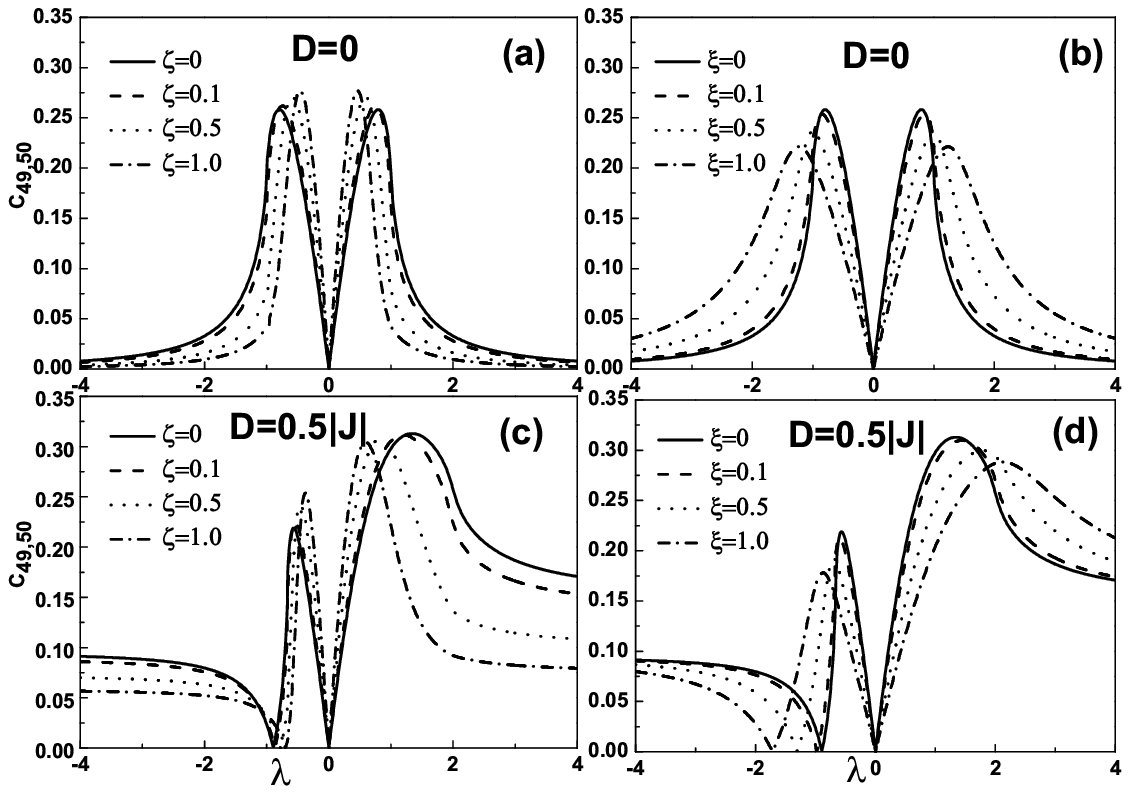}\\
\caption{The nearest neighbour concurrence $C(49,50)$ for
 the impurities $\zeta$ and the impurities $\xi$ as a function of the reduced coupling constant $\lambda$ with
$N =99$, $\gamma=1$, $\kappa=0$.}\label{Fig.1.EPS}
\end{center}
\end{figure}

\begin{figure}
\begin{center}
\includegraphics[width=1.0\textwidth]{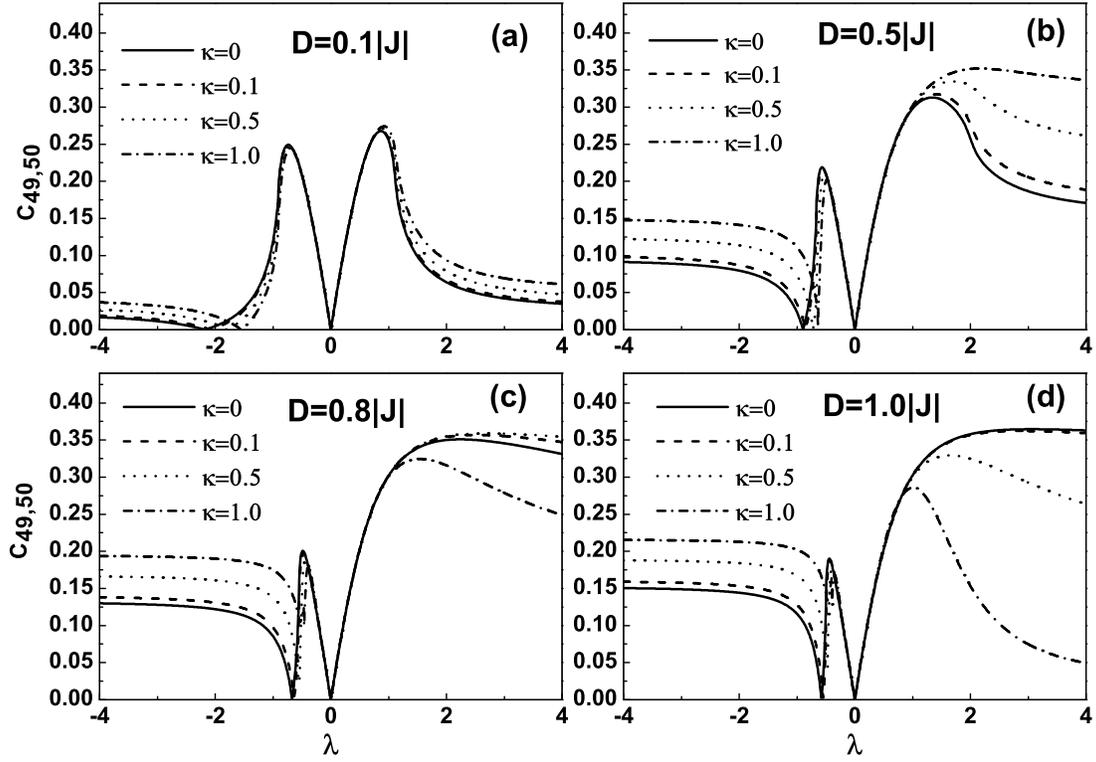}\\
\caption{The nearest neighbour concurrence $C(49,50)$ as a function
of the parameter $\lambda$ for the DM interaction impurities
$\kappa$ with $N=99$, $\gamma=1$, $\zeta=\xi=0$.}\label{Fig.2.EPS}
\end{center}
\end{figure}

\begin{figure}
\begin{center}
\includegraphics[width=1.0\textwidth]{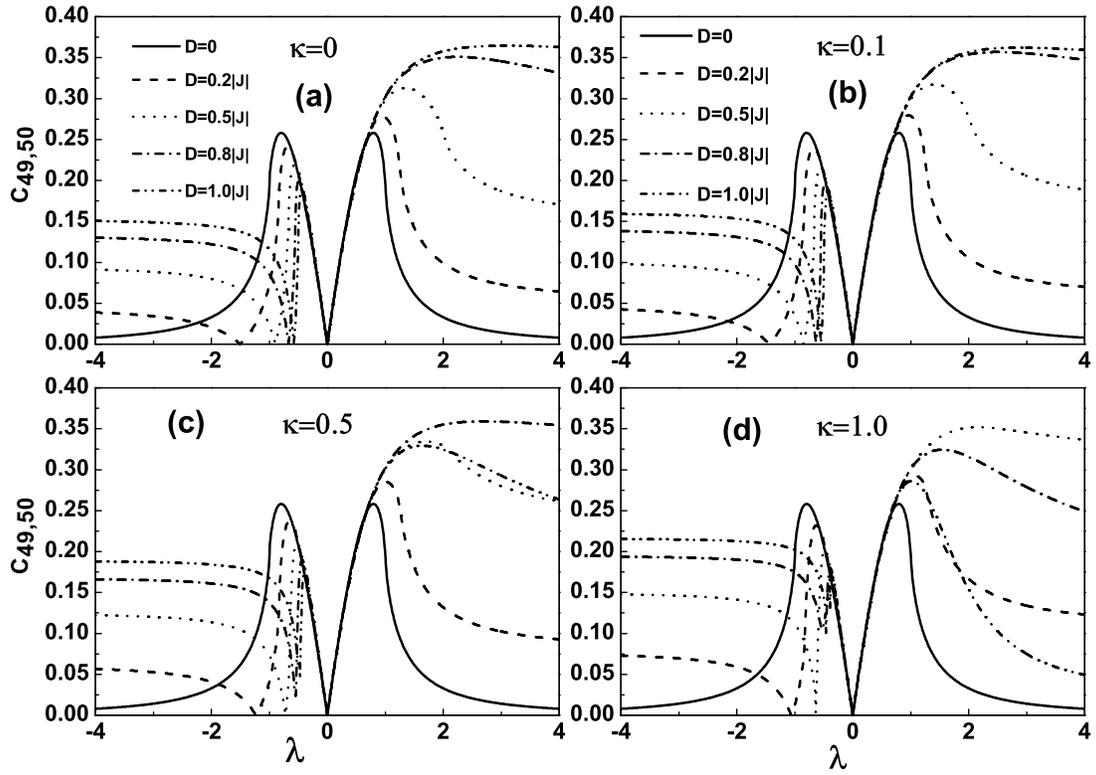}\\
\caption{The nearest neighbour concurrence $C(49,50)$ as a function
of the parameter $\lambda$ for the DM interaction strength $D$ with
$N=99$, $\gamma=1$, $\zeta=\xi=0$.}\label{Fig.3.EPS}
\end{center}
\end{figure}

\end{document}